\documentclass[12pt]{article}
\pdfoutput=1

\usepackage[utf8]{inputenc}
\usepackage[left=2.55cm, right=2.55cm, top=2.55cm, bottom=2.55cm]{geometry}
\usepackage{amsmath,amssymb,amsbsy}
\usepackage{slashed}
\usepackage{xcolor}
\usepackage{graphicx}
\usepackage{url}
\usepackage{cancel}
\usepackage{cite}
\usepackage[colorlinks=true,allcolors=darkpurple,pdfborder={0 0 0},linktocpage=false]{hyperref}
\usepackage{tabularx,booktabs}
\usepackage{multicol}
\usepackage{feynmp}
\usepackage{units}
\usepackage{xspace}
\usepackage[labelfont=bf]{caption}
\usepackage[section]{placeins}
\usepackage{subcaption}
\usepackage{soul} 

\definecolor{darkred}{rgb}{0.6,0,0}
\definecolor{darkpurple}{rgb}{0.5,0,0.5}

\def\hc{\text{h.c.}}

\def\BR{\text{BR}}

\def\z2{$\mathbb{Z}_2$}
\def\321{$\mathrm{SU(3)_c} \times \mathrm{SU(2)_L} \times \mathrm{U(1)_Y}$}
\def\one{\ensuremath{\mathbf{1}}}
\def\two{\ensuremath{\mathbf{2}}}
\def\three{\ensuremath{\mathbf{3}}}
\def\threeS{\ensuremath{\mathbf{\bar 3}}}

\definecolor{vdrgreen}{rgb}{0.0, 0.7, 0.0}
\definecolor{avblue}{rgb}{0.0, 0.0, 0.8}

\newcommand{\AddrIFIC}{%
  Instituto de F\'{i}sica Corpuscular, CSIC-Universitat de Val\`{e}ncia, 46980 Paterna, Spain}

\newcommand{\AddrFISTEO}{%
  Departament de F\'{\i}sica Te\`{o}rica, Universitat de Val\`{e}ncia, 46100 Burjassot, Spain}

\begin{document}

\vspace*{-2cm}
\begin{flushright}
IFIC/22-24 \\
\vspace*{2mm}
\end{flushright}

\begin{center}
\vspace*{15mm}

\vspace{1cm}
{\Large \bf 
Dark matter in the Scotogenic model with spontaneous lepton number violation
} \\
\vspace{1cm}

{\bf Valentina De Romeri$^{\text{a}}$, Jacopo Nava$^{\text{a}}$, Miguel Puerta$^{\text{a}}$, Avelino Vicente$^{\text{a,b}}$}

 \vspace*{.5cm} 
 $^{(\text{a})}$ \AddrIFIC \\\vspace*{.2cm} 
 $^{(\text{b})}$ \AddrFISTEO

 \vspace*{.3cm} 
\href{mailto:deromeri@ific.uv.es}{deromeri@ific.uv.es}, \href{mailto:jacopo.nava@ific.uv.es}{jacopo.nava@ific.uv.es}, \href{mailto:miguel.puerta@ific.uv.es}{miguel.puerta@ific.uv.es}, \href{mailto:avelino.vicente@ific.uv.es}{avelino.vicente@ific.uv.es}
\end{center}

\vspace*{10mm}
\begin{abstract}\noindent\normalsize
Scotogenic models constitute an appealing solution to the generation
of neutrino masses and to the dark matter mystery.  In this work we
consider a version of the Scotogenic model that breaks lepton number
spontaneously.  At this scope, we extend the particle content of the
Scotogenic model with an additional singlet scalar which acquires a
non-zero vacuum expectation value and breaks a global lepton number
symmetry.  As a consequence, a massless Goldstone boson, the majoron,
appears in the particle spectrum.  We discuss how the presence of the
majoron modifies the phenomenology, both in flavor and dark matter
observables. We focus on the fermionic dark matter candidate and 
analyze its relic abundance and prospects for both direct and indirect
detection.
\end{abstract}

\section{Introduction}
\label{sec:intro}

The origin of neutrino masses and the nature of the dark matter (DM) component
of the Universe are two of the most relevant open questions in current
physics. Regarding the former, neutrino oscillation experiments have
robustly established the existence of non-zero neutrino masses and
lepton mixings. In fact, some of the oscillation parameters have been
already determined with great accuracy~\cite{deSalas:2020pgw}. Since
the Standard Model (SM) of particle physics does not include a
mechanism for the generation of neutrino masses, an extension is
called for. Similarly, the Planck collaboration has determined that
about $27 \%$ of the energy-matter content of the Universe is in the form of
DM~\cite{Planck:2018vyg}. It is often assumed that the DM is made of
particles, but no state in the SM spectrum has the required properties
to play such a role. Again, this motivates the exploration of scenarios
beyond the SM.

There are many neutrino mass models. Among them, radiative models
(models that induce neutrino masses at the loop level) are
particularly well motivated, since they naturally explain the
smallness of neutrino masses due to the loop suppression. Pioneer work
on radiative models can be found
in~\cite{Zee:1980ai,Cheng:1980qt,Zee:1985id,Babu:1988ki}, while for a
recent review we refer to~\cite{Cai:2017jrq}. Furthermore, tree-level
contributions to neutrino masses are often forbidden by a conserved
\z2 symmetry which, in addition, stabilizes the lightest \z2-odd
state. Provided it has the correct quantum numbers and can be produced
in the early Universe in the correct amount, this state is a valid DM
candidate. Therefore, radiative models offer a good solution to
simultaneously address the origin of neutrino masses and the DM
problem. A prime example of such class of models is the Scotogenic
model~\cite{Ma:2006km}. This model introduces an additional $\rm
SU(2)_L$ doublet, $\eta$, and three generations of fermion singlets,
$N$, all charged under a \z2 parity. These ingredients suffice to
generate neutrino masses at the 1-loop level and provide a viable DM
candidate.

In most neutrino mass models, neutrinos are Majorana fermions. This is
precisely the case of the Scotogenic model. In this class of models,
$\rm U(1)_L$ --- where $L$ stands for lepton number --- is broken in two
units. The breaking can be explicit, due to the presence of lepton
number violating parameters in the Lagrangian, or spontaneous, if the
minimum of the scalar potential of the model does not preserve the
symmetry. In the \textit{standard} Scotogenic model~\cite{Ma:2006km}
the breaking is explicit. In contrast, in this paper we consider a
version of the Scotogenic model that breaks lepton number
spontaneously. This is achieved by extending the particle content of
the model with an additional singlet scalar, denoted as $\sigma$,
which acquires a non-zero vacuum expectation value (VEV) and breaks
the global $\rm U(1)_L$ symmetry. As a consequence, the
spectrum of the theory contains a massless Goldstone boson, the
majoron,
$J$~\cite{Chikashige:1980qk,Chikashige:1980ui,Gelmini:1980re,Schechter:1981cv,Aulakh:1982yn}. This
state leads to novel phenomenological predictions, both in flavor
observables (due to the existence of new channels such as $\ell_\alpha
\to \ell_\beta \, J$) and in the DM sector (due to the existence of
new processes in the early Universe).

Several works combining spontaneous lepton number breaking with the
Scotogenic generation of fermion masses can be found in the
literature. We highlight~\cite{Bonilla:2019ipe}, which also studies
the DM phenomenology of the Scotogenic model with spontaneous lepton
number violation. We build upon this previous work and go beyond it in
several ways. First of all, our analysis takes into account a wide
variety of lepton flavor violating (LFV) constraints, including
processes that involve the majoron either virtually or as a particle
in the final state. We confirm the results of~\cite{Bonilla:2019ipe},
but also discuss in further detail some aspects of the DM
phenomenology of the model. A high-energy extension of the Scotogenic
model featuring a massless majoron was also introduced
in~\cite{Escribano:2021ymx}, while Ref.~\cite{Ma:2021eko} proposes a
model with spontaneous lepton number violation that induces a small
1-loop mass for a dark Majorana fermion \textit{\`a la Scotogenic}. The
authors of~\cite{Babu:2007sm} studied electroweak baryogenesis in an
extended Scotogenic scenario including a majoron, whereas the possible
Scotogenic origin of the small lepton number violation of the inverse
seesaw was discussed in~\cite{Mandal:2019oth}. Finally, the
spontaneous breaking of a gauged version of lepton number in a
Scotogenic scenario was considered in~\cite{Kang:2021jmi}.

The rest of the manuscript is organized as follows. We present the
model in Sec.~\ref{sec:model}, where we define its basic ingredients,
 discuss its scalar sector and the generation of Majorana
neutrino masses and briefly comment on the possible DM candidates. The most
important experimental bounds that constrain our scenario are
discussed in Sec.~\ref{sec:constraints}, while the results of our
numerical study are presented in Sec.~\ref{sec:results}. Finally, we
summarize and draw our conclusions in Sec.~\ref{sec:conclusions}.

\section{The model}
\label{sec:model}

{
\renewcommand{\arraystretch}{1.6}
\begin{table}[tb]
\centering
\begin{tabular}{ c | c c c c c c | c c c }
\toprule
& $q_L$ & $u_R$ & $d_R$ & $\ell_L$ & $e_R$ & $N$ & $H$ & $\eta$ & $\sigma$ \\ 
\hline
$\rm SU(3)_C$ & $\three$ & $\threeS$ & $\threeS$ & $\one$ & $\one$ & $\one$ & $\one$ & $\one$ &$\one$ \\
$\rm SU(2)_L$ & $\two$ & $\one$ & $\one$ & $\two$ & $\one$ & $\one$ & $\two$ & $\two$ & $\one$ \\
$\rm U(1)_Y$ & $\frac{1}{6}$ & $\frac{2}{3}$ & $-\frac{1}{3}$ & $-\frac{1}{2}$ & $-1$ & $0$ & $\frac{1}{2}$ & $\frac{1}{2}$ & $0$ \\[1mm]
\hline
$\rm U(1)_L$ & $0$ & $0$ & $0$ & $1$ & $1$ & $1$ & $0$ & $0$ & $-2$ \\
\z2 & $+$ & $+$ & $+$ & $+$ & $+$ & $-$ & $+$ & $-$ & $+$ \\
\textsc{Generations} & 3 & 3 & 3 & 3 & 3 & 3 & 1 & 1 & 1 \\
\bottomrule
\end{tabular}
\caption{Particle content of the model and their representations under the gauge and global symmetries. $q_L$, $\ell_L$, $u_R$, $d_R$,
  $e_R$ and $H$ are the usual SM fields.
\label{tab:content}}
\end{table}
}

We consider a variant of the original Scotogenic model. The SM 
particle content is extended by adding the $\rm SU(2)_L$
scalar doublet $\eta$, the scalar singlet $\sigma$ and three
generations of fermion singlets $N$. The scalar doublets of the model
can be decomposed into $\rm SU(2)_L$ components as
\begin{equation}
  H = \begin{pmatrix} H^+ \\ H^0 \end{pmatrix} \, , \quad
  \eta = \begin{pmatrix} \eta^+ \\ \eta^0 \end{pmatrix} \, .
\end{equation}
Here $H$ is the usual SM Higgs doublet. We impose the conservation of
a global $\rm U(1)_L$ symmetry which can be identified with lepton
number. Finally, we also introduce the usual \z2 parity of the
Scotogenic model, under which $N$ and $\eta$ are odd while the rest of
the fields are even.~\footnote{Alternatively, one can assign $\rm
U(1)_L$ charges in such a way that the \z2 parity is obtained as a
remnant symmetry after the spontaneous breaking of lepton
number~\cite{Escribano:2021ymx}. This is more economical in terms of
symmetries, since the usual Scotogenic parity is not imposed, but
automatically obtained from lepton number. However, the generation of
the Scotogenic $\lambda_5$ coupling requires the introduction of a
non-renormalizable operator.} The particle content of the model and
the representations under the gauge and global symmetries are
summarized in Table~\ref{tab:content}.

The most general Yukawa Lagrangian, involving the new particles
compatible with all symmetries, can be written as
\begin{equation} \label{eq:yuk}
    \mathcal{L}_Y = y \, \overline{\ell_L} \, \eta \, N + \kappa \, \sigma \, \overline{N^c} \, N + \hc \, ,
\end{equation}
where $y$ and $\kappa$ are $3 \times 3$ matrices. In the following, we
take $\kappa$ to be diagonal without loss of generality. The most
general scalar potential is given by
\begin{align}
    \mathcal{V} & = m_H^2 \, H^\dagger H + m_\eta^2 \, \eta^\dagger \eta + m_\sigma^2 \, \sigma^* \sigma + \frac{\lambda_1}{2} \, \left( H^\dagger H \right)^2 + \frac{\lambda_2}{2} \, \left( \eta^\dagger \eta \right)^2 + \frac{\lambda_\sigma}{2} \left( \sigma^* \sigma \right)^2 \nonumber \\
    & + \lambda_3 \left( H^\dagger H \right) \left( \eta^\dagger \eta \right) + \lambda_3^{H \sigma} \left( H^\dagger H \right) \left( \sigma^* \sigma \right)  + \lambda_3^{\eta \sigma} \left( \eta^\dagger \eta \right) \left( \sigma^* \sigma \right) \nonumber \\
    & + \lambda_4 \left( H^\dagger \eta \right) \left( \eta^\dagger H \right) + \left[ \frac{\lambda_5}{2} \left(H^\dagger \eta\right)^2 + \hc \right] \, ,
  \label{eq:potential}
\end{align}
where $m_H^2$, $m_\eta^2$ and $m_\sigma^2$ are parameters with dimension
of mass$^2$ and the rest of the parameters are dimensionless.

\subsection{Symmetry breaking and scalar sector}
\label{subsec:sym}

We will assume that the scalar potential parameters are such that a
minimum is found for the configuration
\begin{equation}
  \langle H^0 \rangle = \frac{v}{\sqrt{2}} \, , \quad \langle \eta^0 \rangle = 0 \, , \quad \langle \sigma \rangle = \frac{v_\sigma}{\sqrt{2}} \, .
\end{equation}
Here $v \approx 246$ GeV is the usual electroweak VEV. This vacuum
preserves the \z2 parity, which remains a conserved symmetry. In
contrast, lepton number is spontaneously broken and a Majorana mass
term for the $N$ singlets is induced, with
\begin{equation}
  \frac{M_N}{2} = \kappa \, \frac{v_\sigma}{\sqrt{2}} \, .
\end{equation}
The tadpole equations obtained by minimizing the scalar potential are given by 
\begin{align}
    \frac{\partial \mathcal{V}}{\partial H^0} &= \frac{v}{\sqrt{2}} \left( m_H^2 + \frac{ \lambda_1\, v^2}{2} + \frac{\lambda_3^{H\sigma}\, v_\sigma^2}{2}\right) = 0 \, , \label{eq:min_pot_1} \\
    \frac{\partial \mathcal{V}}{\partial \sigma} &= \frac{v_\sigma}{\sqrt{2}} \left( m_\sigma^2 + \frac{ \lambda_\sigma\, v_\sigma^2}{2} + \frac{\lambda_3^{H\sigma}\, v^2}{2}\right) = 0 \, .\label{eq:min_pot_2}
\end{align}
Assuming the conservation of CP in the scalar sector, one can split
the neutral scalar fields in terms of their real and imaginary
components as
\begin{equation}
  H^0 = \frac{1}{\sqrt{2}} \left( S_H + i \, P_H + v \right) \, , \quad \eta^0 = \frac{1}{\sqrt{2}} \left( \eta_R + i \, \eta_I \right) \, , \quad \sigma = \frac{1}{\sqrt{2}} \left( S_\sigma + i \, P_\sigma + v_\sigma \right) \, .
\end{equation}
The $\eta_R$ and $\eta_I$ fields do not mix with the rest of scalars
due to the \z2 parity. In this case, the scalar potential contains the piece
\begin{equation}
  \mathcal V_{\rm mass}^N = \frac{1}{2} \, \text{Re}(z_i) \, \left( \mathcal{M}_R^2 \right)_{ij} \, \text{Re}(z_j) + \frac{1}{2} \, \text{Im}(z_i) \, \left( \mathcal{M}_I^2 \right)_{ij} \, \text{Im}(z_j) \, ,
\end{equation}
where $z = \{H^0, \sigma \}$ and $\mathcal{M}_R^2$ and
$\mathcal{M}_I^2$ are the $2 \times 2$ CP-even and CP-odd squared mass
matrices, respectively. One finds
\begin{equation} \label{eq:MR2}
  \hspace{-1.2cm} \mathcal{M}_R^2 = \left( \begin{array}{cc}
    m_H^2 + \frac{3\,\lambda_1}{2} v^2 + \frac{\lambda_3^{H\sigma}}{2}v_\sigma^{2}  & \lambda_3^{H\sigma} \, v \, v_\sigma  \\
    \lambda_3^{H\sigma} \, v \, v_\sigma  & m_\sigma^2 + \frac{3 \lambda_\sigma}{2} v_\sigma^{2}+ \frac{\lambda_3^{H\sigma}}{2} v^{2}
    \end{array} \right) \, ,
\end{equation}
and
\begin{equation} \label{eq:MI2}
   \hspace{-1.2cm} \mathcal{M}_I^2 = \left( \begin{array}{cc}
    m_H^2 + \frac{\lambda_1}{2}v^2 + \frac{\lambda_3^{H\sigma}}{2}v_\sigma^{2}  &0  \\
    0  & m_\sigma^2 + \frac{\lambda_\sigma}{2} v_\sigma^{2}+ \frac{\lambda_3^{H\sigma}}{2}v^2  
    \end{array} \right) \, .
\end{equation}
One can now use the tadpole equations in
Eqs.~\eqref{eq:min_pot_1}-\eqref{eq:min_pot_2} to evaluate these
matrices at the minimum of the scalar potential. We obtain
\begin{equation} \label{eq:MR2a}
  \hspace{-1.2cm} \mathcal{M}_R^2 = \left( \begin{array}{cc}
    \lambda_1 \, v^2  & \lambda_3^{H\sigma} \, v \, v_\sigma  \\
    \lambda_3^{H\sigma} \, v \, v_\sigma  &  \lambda_\sigma \,v_\sigma^{2}
    \end{array} \right) \, ,
\end{equation}
while the CP-odd mass matrix becomes identically zero as expected, since it has to provide two massless states: 
the unphysical Goldstone boson $z$ that becomes the longitudinal component of
the $Z$ boson and a physical massless Goldstone boson associated to the spontaneous breaking of the lepton number, the majoron
($J$). Therefore, since $\sigma$ is a gauge singlet field one can make the identification
\begin{equation} \label{CPodd}
J=P_\sigma, \hspace{0.5cm} z=P_H \,.
\end{equation}
The CP-even states $\left\{ S_H, S_\sigma
\right\}$ mix leading to two massive states, $h_1$ and $h_2$ as follows:
\begin{equation} \label{eq:CPeven}
 \left( \begin{array}{c}
    h_1\\
    h_2\end{array} \right) = \mathcal{O} \left( \begin{array}{c}
    S_H\\
    S_{\sigma} \end{array} \right)=  \left( \begin{array}{cc}   
    \cos \alpha & \sin \alpha \\
    -\sin \alpha & \cos \alpha \end{array} \right) \,  \left( \begin{array}{c}
    S_H\\
    S_{\sigma} \end{array} \right) ,
\end{equation}
where $\mathcal{O}$ is the $2\times 2$ orthogonal  matrix which diagonalizes the CP-even mass matrix, such that
\begin{equation}
\mathcal{O} \mathcal{M}_R^2\mathcal{O}^{T}=\text{diag} (m_{h_1}^{2}, m_{h_2}^{2}) \, ,
\end{equation}
and the mass eigenvalues are given by
\begin{equation}
m_{(h_1,h_2)}^2=\frac{\lambda_1}{2}v^2+\frac{\lambda_\sigma}{2}v_\sigma^2 \mp \sqrt{(2\lambda_3^{H\sigma}\, v \,v_\sigma)^{2}+(\lambda_1\,v^2-\lambda_\sigma \, v_\sigma^2)^{2}} \, .
\end{equation}
One of the two scalar masses has to be associated with the $\sim 125$ GeV SM Higgs boson and an additional CP-even state is present in the spectrum.
The angle $\alpha$ is the doublet-singlet mixing angle and is given by 
\begin{equation} \label{eq:Angle}
\tan \,\alpha=\frac{2\lambda_3^{H\sigma}\, v \,v_\sigma}{\lambda_1\,v^2-\lambda_\sigma \, v_\sigma^2-\sqrt{(2\lambda_3^{H\sigma}\, v \,v_\sigma)^{2}+(\lambda_1\,v^2-\lambda_\sigma \, v_\sigma^2)^{2} }} \,.
\end{equation}
We focus now on the \z2-odd scalars. The masses of the CP-even and CP-odd components of $\eta^{0}$ are given by
\begin{equation} \label{eq:NeutralEta}
m^2_{(\eta_R \, , \eta_I)}=m_\eta^2+\frac{\lambda_3^{\eta\sigma}}{2}v_\sigma^{2}+\frac{\lambda_3+\lambda_4\pm \lambda_5}{2}v^2 \, ,
\end{equation}
thus as in the usual Scotogenic model the mass difference between
$\eta_R$ and $\eta_I$ is controlled by the $\lambda_5$
coupling. Finally, the mass of the charged scalar fields $\eta^{\pm}$
turns out to be
\begin{equation}\label{eq:ChargedEta}
m^2_{\eta^\pm}=m_\eta^2+\frac{\lambda_3}{2}v^2+\frac{\lambda_3^{\eta\sigma}}{2}v_\sigma^{2}\, .
\end{equation} 

\subsection{Neutrino masses}
\label{subsec:numass}

\begin{figure}[!t]
  \centering
  \includegraphics[width=0.5\linewidth]{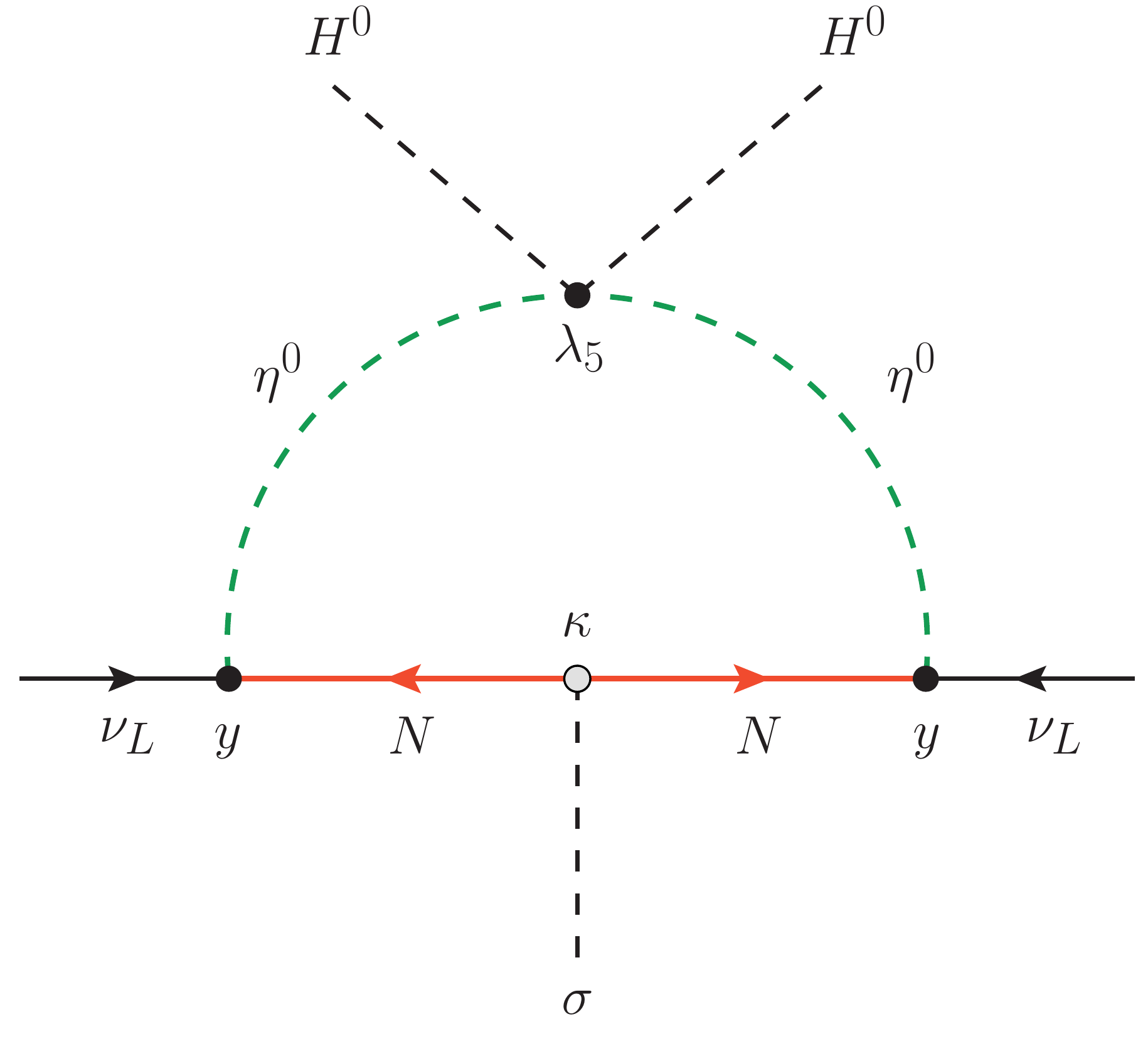}
  \caption{Generation of neutrino masses at the 1-loop level. In this
    diagram, $\eta^0$ denotes the real and imaginary components of the
    neutral component of the $\eta$ doublet.
  \label{fig:numass}}
\end{figure}

Neutrino masses are induced at the 1-loop level, in the same way as in
the standard Scotogenic model, as shown in Fig.~\ref{fig:numass}. One
finds the $3 \times 3$ neutrino mass matrix
\begin{equation} \label{eq:numass}
    (m_\nu)_{\alpha \beta} = \sum_{b=1}^3 \frac{y_{\alpha b} \, y_{\beta b}}{32\pi^2} \, m_{Nb}\left[\frac{m^2_{\eta_R}}{m^2_{Nb}-m^2_{\eta_R}}\log\frac{m^2_{\eta_R}}{m^2_{Nb}}-\frac{m^2_{\eta_I}}{m^2_{Nb}-m^2_{\eta_I}}\log\frac{m^2_{\eta_I}}{m^2_{Nb}}\right] \, ,
\end{equation}
where $m_{\eta_R}$ and $m_{\eta_I}$ are the $\eta_R$ and $\eta_I$
masses, respectively, and $m^2_{Nb}$ are the diagonal elements of
$M_N$. We note that neutrino masses vanish for $m_{\eta_R} =
m_{\eta_I}$. This is consistent with the fact that $m_{\eta_R} -
m_{\eta_I} \propto \lambda_5$, and in the limit $\lambda_5 \to 0$ a
conserved lepton number can be defined. This allows one to assume
$\lambda_5 \ll 1$ in a natural way~\cite{tHooft:1979rat}.

\subsection{Dark matter}
\label{subsec:DM}

The lightest \z2-odd state is completely stable and can, in principle,
be a good DM candidate. In this model, as in the standard Scotogenic
model, this role can be played either by the lightest $N$ state or by
a neutral $\eta$ field ($\eta_R$ or $\eta_I$, depending on the sign of
$\lambda_5$). In this work we will concentrate on the fermion DM and
thus consider $N_1$, the lightest singlet fermion, to be our DM
candidate.

\section{Constraints}
\label{sec:constraints}

Several experimental and theoretical constraints will be considered in
our numerical analysis.

\subsubsection*{Boundedness from below}

We demand the scalar potential to be bounded from below, which implies
the following set of conditions~\cite{Kadastik:2009cu}:
\begin{align}
  \lambda_1,\lambda_2,\lambda_\sigma &\geq 0 \, , \\
  \lambda_3 &\geq - \sqrt{\lambda_1 \, \lambda_2} \, , \\
  \lambda_3^{H \sigma}&\geq-  \sqrt{\lambda_1 \, \lambda_\sigma}\, ,  \label{eq:lambda3Hs} \\
    \lambda_3^{\eta \sigma}&\geq-  \sqrt{\lambda_2 \, \lambda_\sigma}\, , \\
  \lambda_3 + \lambda_4 - |\lambda_5| &\geq - \sqrt{\lambda_1 \, \lambda_2} \, .
\end{align}

\subsubsection*{Higgs boson production and decays}

In our model, all Higgs boson production cross-sections at the LHC are
suppressed with respect to the SM by $c_\alpha^2$, where $c_\alpha =
\cos \alpha$ and $\alpha$ is the mixing angle in the CP-even scalar
sector. In addition, Higgs decays are also affected in two
ways. First, the rates of all visible Higgs decay channels are
universally reduced due the abovementioned mixing. And second, new
decay channels are available. The Higgs boson may decay invisibly to a
pair of majorons or to a pair of DM particles, $h \to J \, J$ and $h
\to N_1 \, N_1$. The former will always be kinematically available,
since the majoron is massless, whereas the latter requires $m_{N_1}
\leq m_h/2$. The CMS collaboration has searched for invisible Higgs
boson decays at the LHC~\cite{CMS:2018yfx}, assuming a completely
SM-like Higgs boson production through vector boson fusion. Therefore
the limit derived in~\cite{CMS:2018yfx} translates into $c_\alpha^2 \,
\BR(h \to \text{invisible}) < 0.19$ at $95\%$ C.L..

A proper phenomenological analysis must take into account both Higgs
production and decays, including visible and invisible ones. In fact,
the recent analysis~\cite{Biekotter:2022ckj} has clearly shown that
the strongest constraints on the parameter space of our model are
obtained by combining the bounds from visible and invisible Higgs
decays. In particular, Figure 9 of this reference shows the limits
obtained for our scenario. These are the constraints that will be
considered in our numerical analysis.

\subsubsection*{Electroweak precision data}

Bounds from electroweak precision data can also be used to contrain
the parameter space of our model. In particular, the oblique
parameters $S$, $T$ and $U$~\cite{Peskin:1991sw} are known to capture
the effect of heavy new fields affecting the gauge boson
propagators. Their current determination is in good agreement with the
SM expectations, although there is some room for new physics. In our
analysis we considered the bounds~\cite{Zyla:2020zbs}
\begin{align}
  S = -0.01 &\pm 0.10 \, , \\
  T = 0.03 &\pm 0.12 \, , \\
  U = 0.02 &\pm 0.11 \, .
\end{align}

\subsubsection*{Neutrino oscillation data}

All the parameter points considered in our analysis comply with the
constraints from neutrino oscillation experiments. This is guaranteed
by means of a modified Casas-Ibarra
parametrization~\cite{Casas:2001sr}, properly adapted to the
Scotogenic
model~\cite{Toma:2013zsa,Cordero-Carrion:2018xre,Cordero-Carrion:2019qtu},
which allows us to express the $y$ Yukawa matrix as
\begin{equation} \label{eq:CI}
  y = \sqrt{\Lambda}^{\: -1} \, R \, \sqrt{\widehat m_\nu} \, U^\dagger \, .
\end{equation}
Here $\Lambda$ is a matrix defined as $\Lambda =
\text{diag}(\Lambda_b)$, with
\begin{equation}
  \Lambda_b = \frac{m_{Nb}}{32\pi^2} \, \left[\frac{m^2_{\eta_R}}{m^2_{Nb}-m^2_{\eta_R}}\log\frac{m^2_{\eta_R}}{m^2_{Nb}}-\frac{m^2_{\eta_I}}{m^2_{Nb}-m^2_{\eta_I}}\log\frac{m^2_{\eta_I}}{m^2_{Nb}}\right] \, ,
\end{equation}
while $R$ is an orthogonal matrix ($R^T R = R R^T = \mathbb{I}$),
generally parametrized by three complex angles. Finally, $U$ is the
unitary matrix that brings $m_\nu$ to diagonal form as $U^T \, m_\nu
\, U = \widehat m_\nu = \text{diag}(m_1,m_2,m_3)$, with $m_i$
  ($i=1,2,3$) the neutrino physical masses. The entries of the unitary
  matrix $U$ as well as the neutrino squared mass differences are
  measured in neutrino oscillation experiments. Our analysis will use
  the results of the global fit~\cite{deSalas:2020pgw}.

\subsubsection*{Majoron diagonal couplings to charged leptons}

The interaction Lagrangian of majorons with charged leptons can be
written as~\cite{Escribano:2020wua}
\begin{equation} \label{eq:llJ}
\mathcal{L}_{\ell \ell J} = J \, \bar{\ell}_\beta \left( S_L^{\beta \alpha} \, P_L + S_R^{\beta \alpha} \, P_R \right)\ell_{\alpha} + \hc \, ,
\end{equation}
where $\ell_{\alpha,\beta}$ are the standard light charged leptons and
$P_{L,R}$ are the usual chiral projectors. The $S_{L,R}$ couplings are
induced at the 1-loop level in our model, as shown
in~\cite{Escribano:2021ymx}. The diagonal $S^{\beta \beta} =
S_L^{\beta \beta} + S_R^{\beta \beta \ast}$ couplings are purely
imaginary, due to the fact that majorons are pseudoscalar states, and
are strongly constrained due to their potential impact on
astrophysical observations. Large couplings to electrons or muons are
excluded since they would lead to an abundant production of majorons
in dense astrophysical media and an efficient cooling
mechanism~\cite{Raffelt:1994ry,DiLuzio:2020wdo,Calibbi:2020jvd,Bollig:2020xdr,Croon:2020lrf}. The
authors of~\cite{Calibbi:2020jvd} used data from white dwarfs to set
the bound
\begin{equation}
  \text{Im} \, S^{e e} < 2.1 \times 10^{-13} \, ,
\end{equation}
while the supernova SN1987A was considered in~\cite{Croon:2020lrf} to
establish the limit~\footnote{Two alternative bounds are given
in~\cite{Croon:2020lrf}. We decided to consider the most conservative
one.}
\begin{equation}
  \text{Im} \, S^{\mu \mu} < 2.1 \times 10^{-9} \, .
\end{equation}
Finally, there are also laboratory bounds on the majoron diagonal
couplings to charged leptons. In~\cite{Escribano:2020wua}, the results
of the OSQAR experiment~\cite{Ballou:2014myz}, a
light-shining-through-a-wall experiment, were used to find the
approximate bounds $S^{ee} \lesssim 10^{-7}$ and $S^{\mu\mu} \lesssim
10^{-5}$. We note that these are clearly less stringent than the
bounds obtained from astrophysical observations.

\subsubsection*{Lepton flavor violation}

As in most neutrino mass models, LFV is a powerful constraint that
strongly restricts the allowed parameter space of our model. Several
processes will be considered in our analysis:

\begin{itemize}
\item The radiative decays $\ell_\alpha \to \ell_\beta \, \gamma$,
  which turn out to be the most constraining ones in most neutrino
  mass models. In particular, the MEG experiment restricts the $\mu
  \to e \gamma$ branching ratio to be smaller than $4.2 \times
  10^{-13}$~\cite{MEG:2016leq}. We also consider the analogous limits
  on $\tau$ LFV decays~\cite{Zyla:2020zbs}, but they are less stringent.
\item The 3-body decays $\ell_\alpha \to \ell_\beta \, \ell_\gamma
  \, \ell_\gamma$, with $\beta = \gamma$ and $\beta \neq \gamma$. In
  this case we follow~\cite{Abada:2014kba} and include the usual
  photon penguin contributions as well as other usually less
  relevant contributions, such as box diagrams. Majoron mediated
  contributions are also included, using the results derived
  in~\cite{Escribano:2020wua}.
\item The decays $\ell_\alpha \to \ell_\beta \, J$ with the majoron in
  the final state are also considered, as they constrain the
  off-diagonal $S_{L,R}$ couplings directly. For instance, the null
  results obtained in the search for the decay $\mu \to e J$ at
  TRIUMF~\cite{Jodidio:1986mz} can be translated into the bound $|S^{e
    \mu}| < 5.3 \times 10^{-11}$~\cite{Hirsch:2009ee}. We use the
  analytical expressions for the majoron off-diagonal couplings to
  charged leptons in the Scotogenic model found
  in~\cite{Escribano:2021ymx}.
  
\item $\mu-e$ conversion in nuclei, again
  following the analytical results in~\cite{Abada:2014kba}.
\end{itemize}

\subsubsection*{Majoron coupling to neutrinos}
  
Constraints on the couplings of the majoron to neutrinos can be
derived from astrophysics, since the majoron can have a significant
impact on the explosion and cooling of supernovae (see
e.g.~\cite{Farzan:2002wx}) and also from cosmic microwave background
data~\cite{Forastieri:2015paa}.  Laboratory experiments searching for
neutrinoless double beta decays (e.g.~\cite{Berezhiani:1992cd}) and
possible effects on meson and lepton decays~\cite{Lessa:2007up} also
set limits on the magnitude of neutrino-majoron couplings. Among
these, the most stringent ones are those derived from astrophysics,
with constraints in the $\sim 10^{-7}$ ballpark. In our model, the
interaction of majorons with neutrinos arises at the 1-loop level,
similarly to the interaction with charged leptons, and the
corresponding couplings are thus expected to be of the same
order. Since the constraints on the couplings to charged leptons are
orders of magnitude more stringent, we can safely ignore the
constraints on the couplings to neutrinos in our analysis.

\section{Numerical results}
\label{sec:results}

 We now proceed to discuss the results of our analysis.  To perform
 the numerical scan, we have first implemented the model in
 \texttt{SARAH (version 4.11.0)}~\cite{Staub:2013tta}, a {\tt
   Mathematica} package for the analytical evaluation of all the
 information about the model.~\footnote{See~\cite{Vicente:2015zba} for
 a pedagogical introduction to the use of {\tt SARAH}.} With this
 tool, we have created a source code for \texttt{SPheno (version
   4.0.2)}~\cite{Porod:2003um,Porod:2011nf}, thus allowing for an
 efficient numerical evaluation of all the analytical expressions
 derived with \texttt{SARAH}. We have also computed several
 observables of interest in our model, including the lepton flavor
 violating ones, both analytically and with the help of {\tt
   FlavorKit}~\cite{Porod:2014xia}, for an in-depth cross-check of
 their expressions. Finally, we have used \texttt{micrOmegas
   (version 5.0.9)}~\cite{Belanger:2018ccd} to obtain the main
 DM observables, namely the DM relic density and direct and indirect
 detection predictions.

\begin{table}[t]
    \centering
    \begin{tabular}{|c|}
    \hline
         $\lambda_{2,3,4,\sigma} \in \left[10^{-6}, 1 \right]$ \\
         $\lambda_{5} \in \left[10^{-8}, 1 \right]$\\
         $m_{h_2} \in \left[20, 2000 \right]$ GeV\\
         $\kappa_{11} \in \left[0.01, 1 \right]$\\
         $m_\eta^2 \in \left[10^5, 10^7 \right]$ GeV$^2$ (or fixed) \\
         $v_\sigma  \in \left[0.5, 10 \right]$ TeV \\
        \hline
    \end{tabular}
    \caption{Values of the main input parameters for the numerical scan. 
    \label{tab:inputparameters}}
\end{table}

As already mentioned, while both the scalar $\eta_{R,I}$ and the
fermions $N_i$ are, in principle, viable DM candidates in this model,
in our analysis we focus on the lightest Majorana fermion $N_1$ as the
main component of the DM.  We summarize our choice of parameters for
the numerical scan in Tab.~\ref{tab:inputparameters}. Moreover,
$\lambda_1$ is fixed by the condition of requiring $m_{h_1} = 125$
GeV. In some numerical scans we have fixed the value of $m_{h_2} =
500$ GeV or the value of $m_\eta^2$ such that $m_{\eta_{R,I},
  \eta^\pm} - m_{N_1} \lesssim 20$ GeV, as we will discuss in more
detail below. With our choice of parameters, all the parameter points
considered in our numerical scans easily pass the bounds from the
$S,T,U$ parameters. We note that with our choice of $\lambda_4$ and
$\lambda_5$ values the mass splitting between the neutral and charged
components of the $\eta$ doublet is small (see
Eqs.~\eqref{eq:NeutralEta} and \eqref{eq:ChargedEta}). We have
explicitly checked that only for $\lambda_4 \gtrsim 2$ the electroweak
precision bounds become relevant, but we did not explore this region
of parameter space in our scans.  Since we want to focus on $N_1$ as
the DM candidate, we further require that $m_{N_1} < m_{N_{2,3}},
m_{\eta_{R,I}}$. We have chosen normal hierarchy for the neutrino
spectrum and considered the best-fit values for the neutrino
oscillation parameters found by the global
fit~\cite{deSalas:2020pgw}. Finally, the three angles in the
orthogonal $R$ matrix are assumed to be real and taken randomly in our
numerical scans.~\footnote{While more general scans with complex $R$
matrices are in principle possible, we expect little impact on the DM
phenomenology discussed here. Only in some specific regions of
parameter space one may expect a change, due to the occurrence of very
large Yukawa couplings at the prize of large
cancellations~\cite{AristizabalSierra:2011mn}, which we consider
tuned.}\\

\begin{figure}[!hbt]
  \centering
  \includegraphics[width=0.8\linewidth]{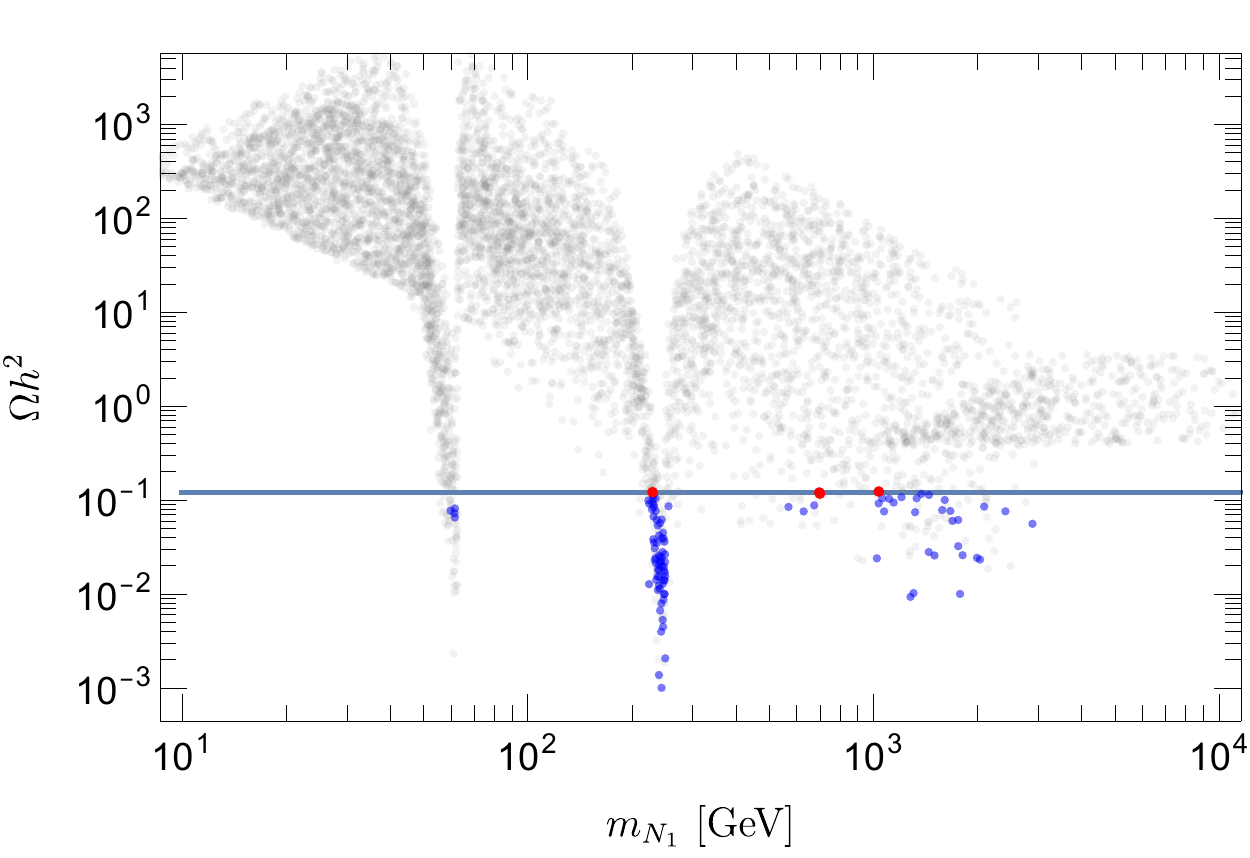}
  \caption{Relic abundance of $N_1$ as a function of $m_{N_1}$. Red points depict solutions in agreement with the cold DM measurement obtained from Planck data~\cite{Planck:2018vyg} (the blue thin band shows the 3$\sigma$ interval) while blue points depict solutions leading to underabundant DM. Gray points are excluded by any of the constraints listed in Sec.~\ref{sec:constraints} or due to an overabundant DM relic density.  }
  \label{fig:omega-mN1}
\end{figure}

We first show in Fig.~\ref{fig:omega-mN1} the relic abundance of
$N_1$, as a function of its mass. For this specific scan, we have
fixed $m_{h_2} = 500$ GeV to highlight the $s$-channel annihilation of
$N_1$ via $h_2$.  In this figure, grey points denote solutions either
leading to overabundant DM or excluded by any of the constraints
listed in Sec.~\ref{sec:constraints}, or where the spin-independent
$N_1$-nucleon elastic scattering cross section is excluded by the most
recent data from the LUX-ZEPLIN experiment~\cite{LZ:2022ufs}. Red
points denote solutions which can reproduce the observed cold DM relic
density, as they fall within the 3$\sigma$ range obtained by the
Planck satellite data~\cite{Planck:2018vyg}, $\Omega_{N_1} h^2 =
0.120 \pm 0.0036$ (blue thin band). Solutions leading to underabundant
DM (which would then require another DM candidate to explain the
totality of the observed cold DM relic density) are depicted in
blue. As can be seen from the plot, most of the solutions lead to
overabundant DM, except for points falling in the following regions:
(i) a resonant region where $m_{N_1} \sim m_{h_1}/2 \sim 60$ GeV, (ii) a
second resonant region where $m_{N_1} \sim m_{h_2}/2 \sim 250$ GeV and
(iii) a region of coannihilations at higher $m_{N_1}$.
 
\begin{figure}[!hbt]
  \centering
  \includegraphics[width=0.8\linewidth]{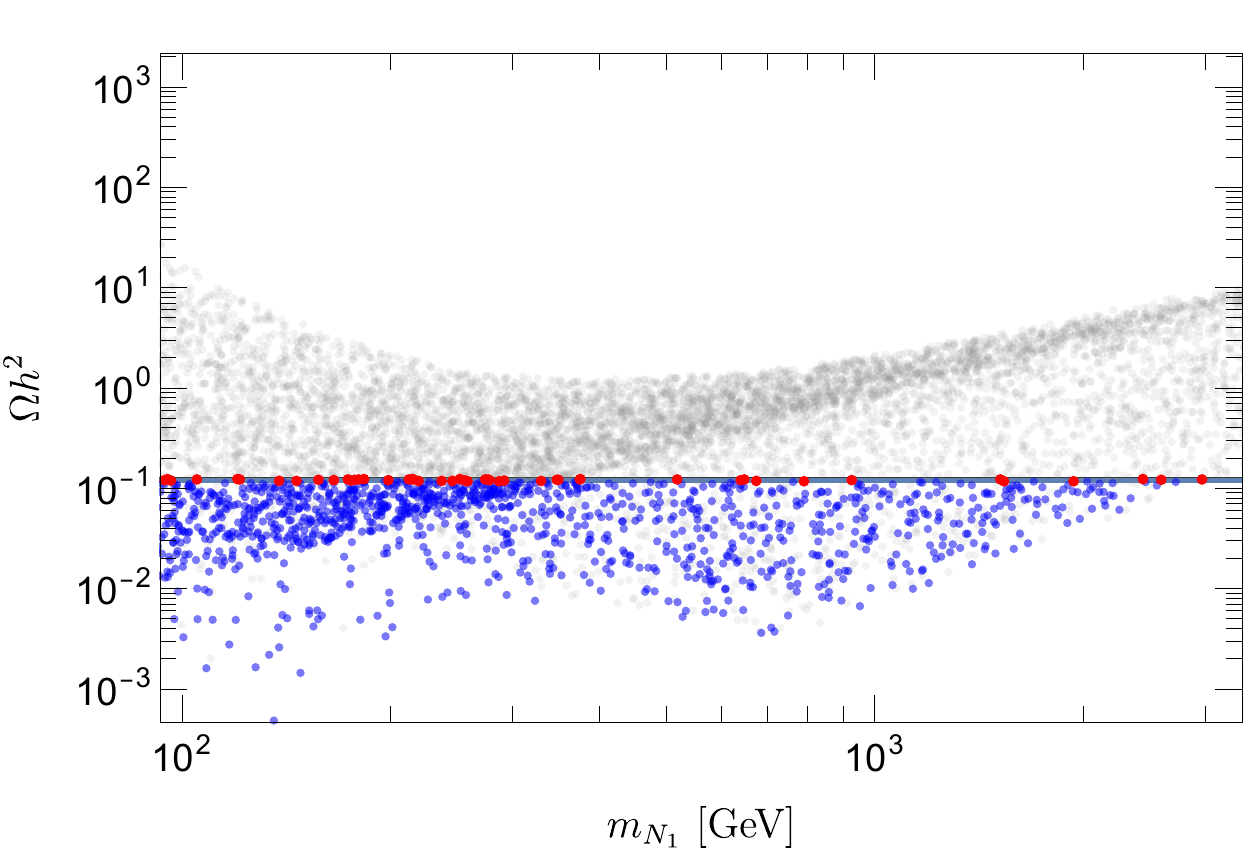}
  \caption{Relic abundance of $N_1$ as a function of $m_{N_1}$ in the coannihilation region, where $\Delta \in \left[0, 20\right]$ GeV. Same color code as in Fig.~\ref{fig:omega-mN1}. 
  \label{fig:omega-mN1-co}}
\end{figure}

To explore in more detail the third, high-mass region, we performed a
second numerical scan in which we have varied the mass difference
$\Delta = m_{\eta_R} - m_{N_1}$ in the $\left[ 0, 20\right]$ GeV
range. In such a way, we have enforced $N_1$ to be in the $\sim [100 -
  3000]$ GeV region, where coannihilations with $\eta_{R,I}$ and
$\eta^\pm$ are very relevant, thus reducing the relic abundance of
$N_1$.  Figure~\ref{fig:omega-mN1-co} shows this region in parameter
space, in which the DM relic density is set by coannihilations.  The
color code is the same as in Fig.~\ref{fig:omega-mN1}. Compared to the
result of the previous scan (Fig.~\ref{fig:omega-mN1}), we can see
that if coannihilations are relevant, more viable solutions can be
found in the $m_{N_1} \sim \left[100, 3000 \right]$ GeV region. We
clarify that $\Delta < 20$ GeV is not motivated by any symmetry
argument, but just a convenient parameter choice to focus our
numerical analysis on a region in which coannihilations are more
effective. Finally, we should also make a comment about the region
  with light $\eta$ states ($m_{\eta_{R,I}}, m_{\eta^\pm} \lesssim
  250$ GeV). These states can be pair-produced at the LHC via
  Drell-Yan processes. However, our choice of $\Delta$ implies a
  compressed spectrum with $N$, $\eta_{R,I}$ and $\eta^\pm$ in a
  narrow window of just $20$ GeV, thus implying soft leptons in the
  $\eta \to N \ell$ final decay. Searches for this type of signal
  exist, although not dedicated to our specific scenario. For
  instance, the ATLAS collaboration looked for direct slepton
  production with a compressed spectrum in~\cite{ATLAS:2019lng}. This
  analysis assumes mass-degenerate 1st and 2nd generation sleptons
  decaying to flavor conserving final states, whereas our scenario
  contains only a copy of the $\eta$ doublet (and not two), with both
  flavor-conserving and flavor-violating decays. Therefore, the
  obtained limits are not applicable. Nevertheless, we note that some
  points with $\Delta \sim 10$ GeV, where the experimental searches
  are more efficient, must be excluded. A detailed analysis including
  this constraint is clearly beyond the scope of our work and would
  not have any impact on our conclusions. \\

\begin{figure}[!hbt]
  \centering
  \includegraphics[width=0.8\linewidth]{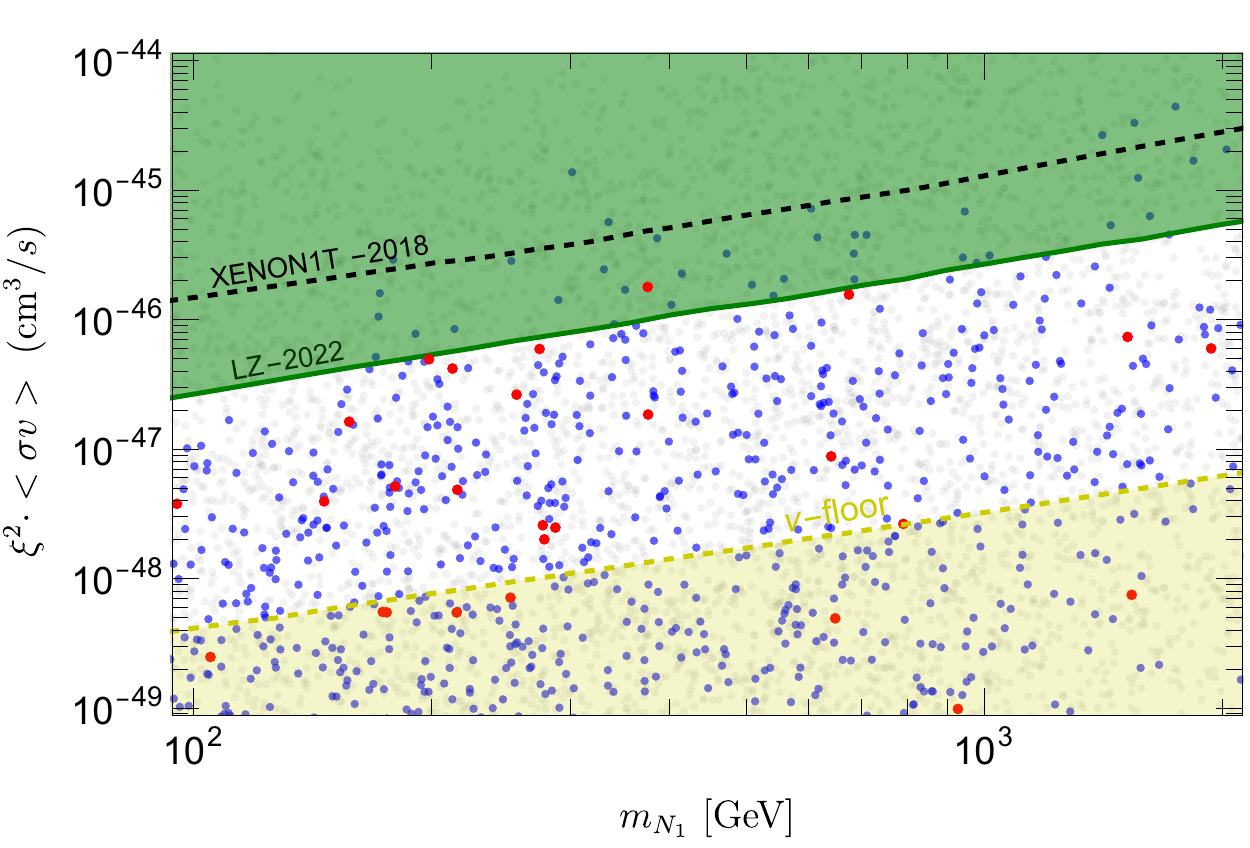}
  \caption{Spin-independent $N_1$-nucleon elastic scattering cross
    section -- weighted by the relative abundance -- as a function of
    $m_{N_1}$.  The green area is already excluded by the LUX-ZEPLIN
    experiment (LZ-2022)~\cite{LZ:2022ufs}, while the black dashed
    line denotes the constraint from XENON1T
    (XENON1T-2018)~\cite{Aprile:2018dbl}. The dashed orange curve
    indicates the expected discovery limit corresponding to the
    $\nu$-floor from CE$\nu$NS of solar and atmospheric neutrinos for
    a Ge target~\cite{Billard:2013qya}.
  \label{fig:DD}}
\end{figure}

Next we discuss the results for $N_1$ direct detection. In order to
maximize the number of viable solutions, we focus again on the
coannihilation region, and we show in Fig.~\ref{fig:DD} the
spin-independent $N_1$-nucleon elastic scattering cross section,
$\sigma_{\rm SI}$, as a function of the DM mass, $m_{N_1}$. The cross
section shown in this figure is weighted by the relative abundance
$\xi$, defined as
\begin{equation}
  \xi = \frac{\Omega_{N_1}}{\Omega_{\text{\rm DM,Planck}}} \, ,
\end{equation}
where $\Omega_{\rm DM,Planck} h^2 = 0.120$~\cite{Planck:2018vyg}. We
apply the same color code as in Fig.~\ref{fig:omega-mN1}, that is red
points indicate solutions explaining the totality of observed DM,
while blue points denote under-abundant DM. The plain green line and
dashed area indicate the current most stringent limit from the
LUX-ZEPLIN experiment (LZ-2022)~\cite{LZ:2022ufs}, while the black
dashed line denotes the constraint from XENON1T
(XENON1T-2018)~\cite{Aprile:2018dbl}. Other (less stringent)
constraints on $\sigma_{\rm SI}$ apply from the liquid xenon
experiment PandaX-II~\cite{Cui:2017nnn} and from liquid argon
experiments like DarkSide-50~\cite{Agnes:2018fwg} and
DEAP-3600~\cite{Ajaj:2019imk}, although they are not shown
here. Future facilities including XENONnT~\cite{Aprile:2020vtw},
DarkSide-20k~\cite{DS_ESPP}, ARGO~\cite{DS_ESPP} and
DARWIN~\cite{Schumann:2015cpa,Aalbers:2016jon}
(see~\cite{Billard:2021uyg} for an overview) will be able to further
inspect the parameter space of this model. As for general reference,
we further illustrate the expected discovery limit corresponding to
the so-called ``$\nu$-floor" from coherent elastic neutrino-nucleus
scattering (CE$\nu$NS) for a Ge target~\cite{Billard:2013qya} (dashed
orange line).~\footnote{Notice, however, that this should not be taken
as a hard limit, as it can be overcome with different techniques and
it has strong dependences on both the target material and a series of
uncertainties (see for
example~\cite{AristizabalSierra:2021kht,OHare:2021utq,Akerib:2022ort}
for more details).}

\begin{figure}[!hbt]
  \centering
  \includegraphics[width=0.8\linewidth]{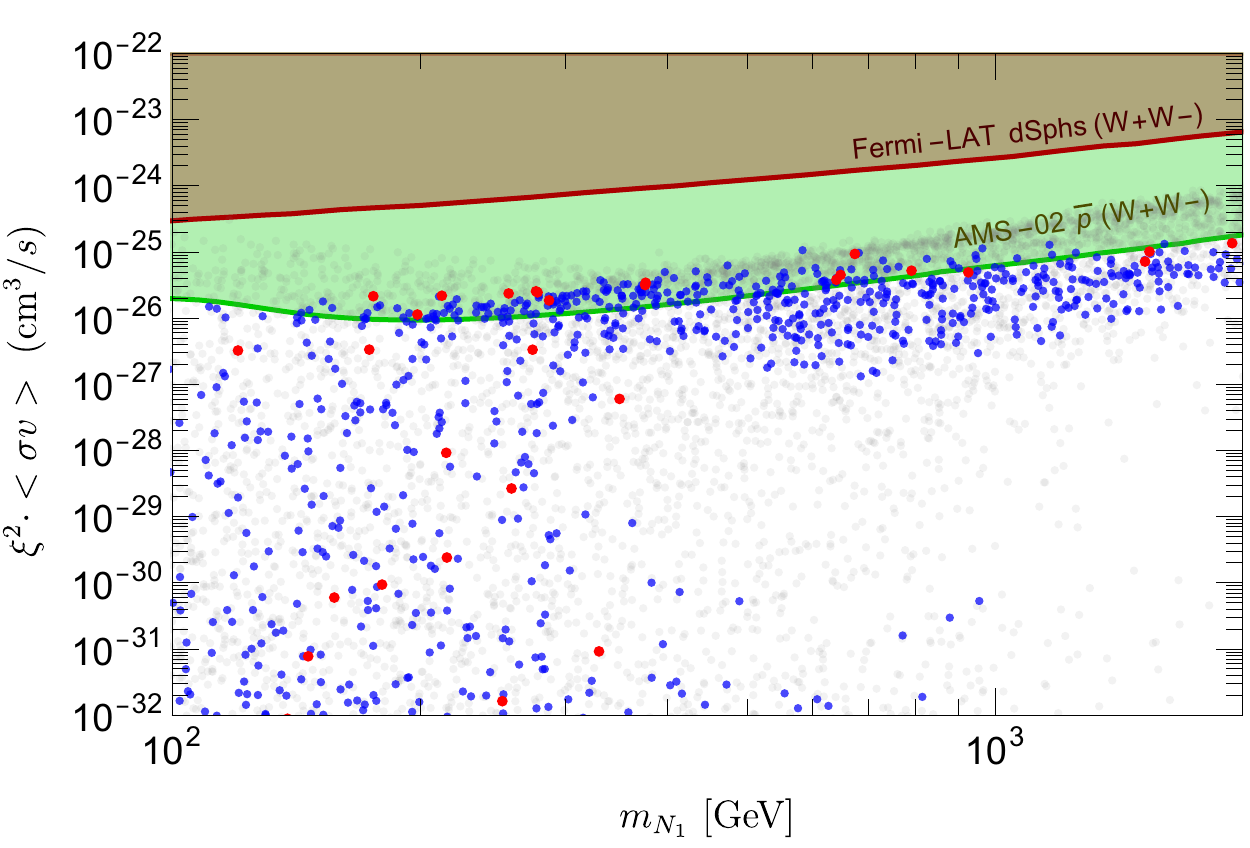}
  \caption{$N_1$ total annihilation cross section as a function of $m_{N_1}$. The red and green lines refer to the corresponding 95\% C.L. upper limits currently set by Fermi-LAT gamma-ray data from dSphs~\cite{Ackermann:2015zua} and from the antiproton and B/C data of AMS-02~\cite{Reinert:2017aga}, respectively. \label{fig:ID}}
\end{figure}

Finally, we have explored the predictions for the velocity-averaged
cross section of $N_1$ annihilation into gamma rays. These are among
the most suitable messengers to probe DM via indirect detection. We
focus once more on the coannihilation region, i.e. on the high-mass
range $m_{N_1} \sim 0.1 - 2$ TeV, where the annihilation channels $N_1
N_1 \rightarrow h_1 h_1, h_2 h_2, h_1 h_2, Z^0 Z^0, h_i J$ can be
relevant. The hadronization of the final-state gauge bosons and Higgs
bosons will produce neutral pions, which in turn can decay into
photons thus giving rise to a gamma-ray flux with a continuum
spectrum which may be within reach of DM indirect detection
experiments. While a detailed calculation of the gamma-ray energy
spectra produced by the annihilation of two $N_1$ particles in this
specific model should be performed, in order to correctly compute
exclusion bounds from existing gamma-ray data, this is out of the
scope of this work. However, we can notice that the main annihilation
channels in this high-mass range include Higgs bosons in the final
state. The gamma-ray energy spectrum from DM DM $ \rightarrow h_1 h_1$
annihilation channel is very similar to that from DM DM $\rightarrow
W^+ W^-$ at $m_{N_1} \sim 1$ TeV (see for instance Fig. 15
of~\cite{Cirelli:2010xx}). In the following, for the sake of
simplicity, we will compare our predictions with bounds obtained
assuming $W^+ W^-$ as the main annihilation channel, to get an overall
idea of how current data can constrain the parameter space of this
model.\\ Charged cosmic rays can also be used to look for $N_1$
annihilations, even though their detection is more challenging due to
uncertainties in the treatment of their propagation. For instance,
AMS-02 data on the antiproton flux and the Boron to Carbon (B/C) ratio
can be used to constrain the $N_1$ annihilation cross
section~\cite{Aguilar:2016kjl,Reinert:2017aga,Cuoco:2017iax}. With
some caveats concerning the astrophysical uncertainties on the
$\bar{p}$ production, propagation and on solar modulation (see
e.g.~\cite{Cuoco:2019kuu,Cholis:2019ejx,Heisig:2020nse}), these bounds
turn out to be stronger than gamma-ray limits from dwarf spheroidal
satellite galaxies in some mass ranges. Following the same
considerations as before, i.e. that the antiproton energy spectrum
from DM DM $ \rightarrow h_1 h_1$ annihilation channel is very similar
to that from DM DM $\rightarrow W^+ W^-$ at $m_{N_1} \sim 1$ TeV, we
will compare our predictions to current limits on the $N_1$
annihilation cross section set by combination of $\bar{p}$ and B/C
data of AMS-02~\cite{Aguilar:2016kjl,Reinert:2017aga} assuming $W^+
W^-$ as the dominant annihilation channel.  We show in
Fig.~\ref{fig:ID} the $N_1$ total annihilation cross section ---
weighted by $\xi^2$ --- versus its mass. The color code follow the
same scheme as in Figs.~\ref{fig:omega-mN1-co},~\ref{fig:DD}.  We also
depict the 95\% C.L. upper limits currently set by the Fermi-LAT with
gamma-ray observations of Milky Way dSphs (6 years, Pass 8 event-level
analysis)~\cite{Ackermann:2015zua} (red solid curve and shaded area)
and from a combination of $\bar{p}$ and B/C data of
AMS-02~\cite{Aguilar:2016kjl,Reinert:2017aga} (green), both assuming
$N_1 N_1 \rightarrow W^+ W^-$ as main annihilation channel due to the
considerations made before.  We see that few solutions already fall
within the region currently excluded by AMS-02 data. As already
highlighted, while a dedicated analysis should be performed for this
specific model, we can conclude that current $\bar{p}$ and B/C data
may be already excluding a relevant part of the parameter
space. Forthcoming data will allow to further probe $N_1$ as a DM
candidate via its multi-messenger signals.\\

\begin{figure}[!hbt]
  \centering
  \includegraphics[width=0.8\linewidth]{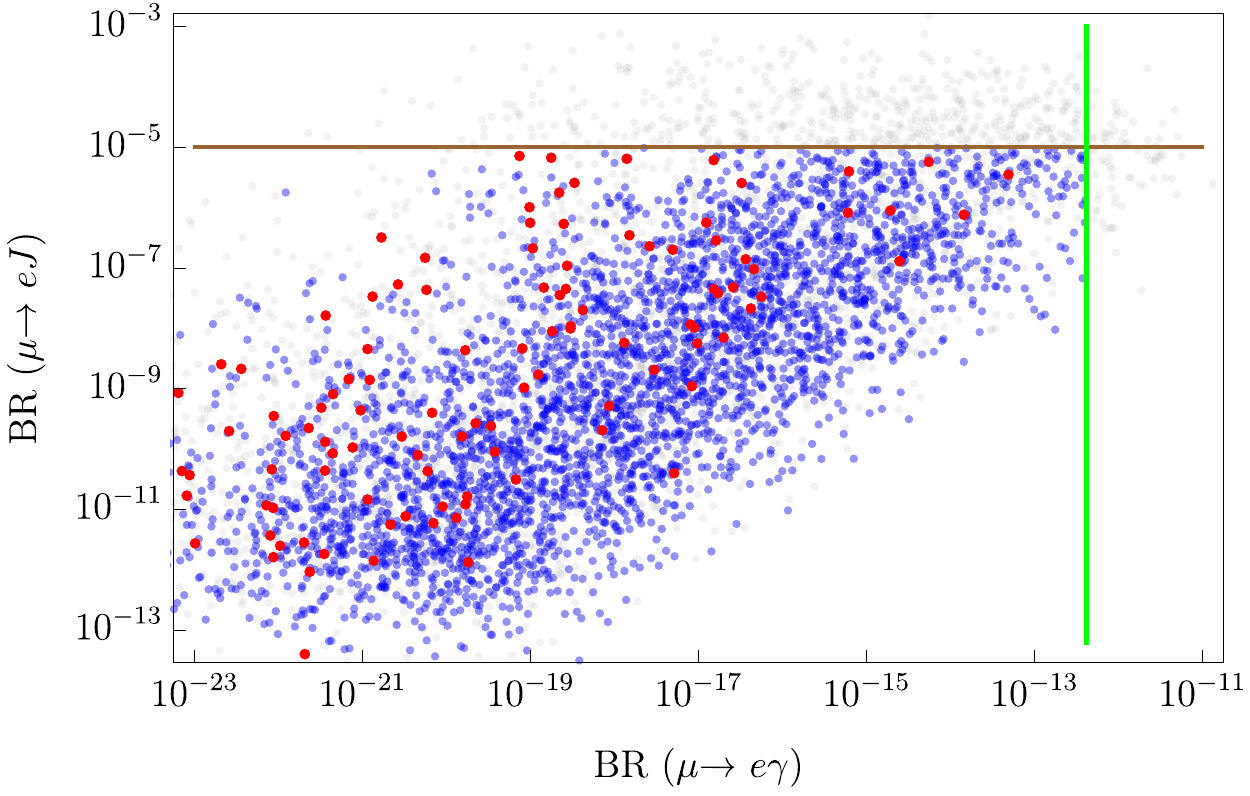}
  \caption{BR($\mu \to e J$) as a function of BR($\mu \to e \gamma$)
    in the coannihilation region, where $\Delta \in \left[0,
      20\right]$ GeV. Same color code as in
    Fig.~\ref{fig:omega-mN1}. The horizontal and vertical lines
    correspond to the current experimental limits, discussed in
    Sec.~\ref{sec:constraints}.
  \label{fig:BRsLFV}}
\end{figure}

As in many scenarios for neutrino mass generation, LFV processes
strongly restrict the available parameter space of the model. In
addition to $\mu \to e \gamma$, very commonly considered in
phenomenological studies, our model also leads to signatures with the
majoron in the final state, like $\mu \to e
J$. Figure~\ref{fig:BRsLFV} shows BR($\mu \to e J$) as a function of
BR($\mu \to e \gamma$). Again, we have focused on the coannihilation
region. We first notice that some parameter points are already
excluded by the current experimental limits on these LFV branching
ratios. However, one can also see that our numerical scan also finds
many valid parameter points leading to very low values of both BR($\mu
\to e \gamma$) and BR($\mu \to e J$), clearly below the discovery
reach of planned experiments. This is not surprising, since we take
random $R$ matrices in our numerical scans, hence accidentally finding
parameter points with suppressed $\mu-e$ flavor violation. While a
slight correlation among these two observables can be observed in
Fig.~\ref{fig:BRsLFV}, $\mu \to e \gamma$ receives contributions from
additional loop diagrams that do not involve the majoron. The two
observables are hence independent. Interestingly, we find that BR($\mu
\to e J$) is generally more constraining than BR($\mu \to e \gamma$),
although the difference is not very significant.

\section{Summary and discussion}
\label{sec:conclusions}

Most SM extensions aiming at an explanation of neutrino oscillation
data consider Majorana neutrinos. This option breaks the accidental
$\rm U(1)_L$ lepton number symmetry of the SM in two units. If the
breaking of lepton number is spontaneous, a Goldstone boson appears in
the particle spectrum of the theory, the majoron. In this work we have
analyzed the dark matter phenomenology of this scenario in the context
of the popular Scotogenic model.\\
Focusing on the fermionic DM
candidate $N_1$, we have found that it can explain the observed DM
abundance in three regions of parameter space: (i) a resonant region
where it annihilates via $h_1$, with $m_{N_1} \sim 60$ GeV, (ii) a
second resonant region where $s-$channel annihilations via $h_2$ are
relevant and (iii) a region of coannihilations at $m_{N_1} \sim 1$
TeV. In particular, if coannihilations are relevant, more allowed
solutions are found, either explaining the totality of DM or at least
a sizeable part of it.  While some of these solutions are already
excluded by the recent LUX-ZEPLIN result, most of them are within the
reach of near-future direct detection experiments.  Interestingly,
indirect detection searches seem to constitute another promising tool
to further probe $N_1$ as a DM candidate via its multi-messenger
signals, mainly gamma rays and antiprotons.  All in all, the presence
of the majoron and of a second Higgs open up the allowed parameter
space of $N_1$ as DM, compared to the standard Scotogenic model.  The
majoron has an impact also on the phenomenology of LFV observables, as
it leads to new interesting signatures, where it appears in the final
state.  Among these, we found that BR($\mu \to e J$) is generally more
constraining than the most common BR($\mu \to e \gamma$).\\
Moreover, let us comment
that the presence of a massless majoron may have relevant implications
on the early-Universe cosmology. In particular, it can affect
cosmological and astrophysical environments, and can contribute to
$\Delta N_\mathrm{eff}$.
In principle, these bounds could be relaxed if the majoron acquires a small mass (for instance from quantum gravity considerations).
In such a case, the majoron would decay before Big Bang nucleosynthesis and would not affect cosmological observations. 
However, let us notice that, in order not to alter the phenomenological analysis presented in this paper, the majoron mass should be smaller than the electron one and hence its only available decay channel would be into active neutrinos. On the other hand, the majoron can be produced from the Higgs
decay, or the annihilation of $N_1$ or even via freeze-in through its
 small coupling with the active neutrinos.  
 If it is massless and thermalizes, in order to avoid
constraints from $\Delta N_\mathrm{eff}$, one should require 
the majoron to decouple before $T \sim 0.5$ GeV (see e.g. \cite{Baumann:2016wac}) to avoid the current constraint from Planck. This can be easily obtained if $\lambda_3^{H\sigma}$ is set small enough ($\lesssim 10^{-5}$). In such a case, all majoron production channels through SM particles would be suppressed, and it could only be produced via interactions with $N_1$, through the (sizeable) coupling $\kappa$. If this is the case, the majoron would freeze out at around the same time as $N_1$, that is at $T_F \sim m_{N_1}/20$, thus not substantially contributing to $\Delta N_\mathrm{eff}$.
We have checked that by imposing $\lambda_3^{H\sigma} \lesssim 10^{-5}$ our results remain almost unchanged, with the only exception of the second resonance shown in Fig.~\ref{fig:omega-mN1}. This region would disappear, due to the fact that a tiny $\lambda_3^{H\sigma}$ suppresses all vertices involving a (heavy or light) Higgs.

Finally, another interesting scenario consists in $N_1$
having very tiny couplings, so that it does not reach thermal
equilibrium in the early Universe and it is instead produced via
freeze-in.  Such a production mechanism, yet together with the
presence of the majoron, should also lead to some interesting
phenomenology. We leave such analysis for a follow-up of this paper.

\section*{Acknowledgements}

The authors are grateful to V\'ictor Mart\'in-Lozano for enlightening
discussions on collider constraints in our scenario. Work supported by
the Spanish grants PID2020-113775GB-I00 (AEI/10.13039/501100011033),
CIPROM/2021/054, SEJI/2018/033 and SEJI/2020/016 (Generalitat
Valenciana).
AV acknowledges financial support from MINECO through the Ramón y
Cajal contract RYC2018-025795-I. VDR acknowledges financial support by
the Universitat de Val\`encia through the sub-programme “ATRACCI\'O DE
TALENT 2019”, in the early stages of this work.



\bibliographystyle{utphys}
\bibliography{refs}

\end{document}